\documentclass[prl,twocolumn,showpacs,amsmath,amssymb]{revtex4}

\usepackage{graphicx,dcolumn}% Align table columns on decimal point
\usepackage{bm}% bold math

\begin{document}

\title{Ensemble averageability in network spectra}

\author{Dong-Hee Kim} 
\author{Adilson E. Motter} 
\affiliation{Department of Physics and Astronomy, Northwestern University, 
Evanston, Illinois 60208, USA}

\date{\today}

\begin{abstract}
The extreme eigenvalues of connectivity matrices govern the
influence of the network structure on a number of 
network dynamical processes.
A fundamental open question is whether the eigenvalues of large networks are well
represented by ensemble averages.
Here we investigate this question explicitly and validate the concept of 
{\it ensemble averageability} in random scale-free networks by showing that
the ensemble distributions of extreme eigenvalues converge to peaked distributions as
the system size increases. We discuss the significance of this result 
using synchronization and epidemic spreading as example processes. 
\end{abstract}
\pacs{05.50.+q, 05.10.-a, 87.18.Sn, 89.75.-k}

\maketitle

The structure and dynamics of complex networks 
is of increasing interest in nonlinear dynamics, biological 
physics, complex systems, and statistical physics~\cite{Motter2006,nbw:2006,b:2006}. 
Part of this interest comes from the realization that commonly observed 
structural properties, such as the scale-free (SF) degree 
distribution~\cite{Albert2002}, strongly influence 
the collective dynamics of the system. 
In many dynamical processes, the influence of the network structure is 
encoded in the extreme eigenvalues of a connectivity matrix. 
In complete synchronization and consensus phenomena, for example, 
the stability and convergence are often determined by the largest and smallest
nonzero eigenvalues of the Laplacian 
matrix~\cite{Barahona2002,Nishikawa2003,consensus}.
In diffusion processes, the relaxation rate is governed
by the corresponding eigenvalues of the normalized Laplacian~\cite{Motter2005}.
The largest eigenvalue of the adjacency matrix, on the other hand, 
plays a central role in determining epidemic thresholds~\cite{WCWF,Boguna2003} 
and critical couplings for the onset of coherent behavior~\cite{Restrepo2006}. 

Our ultimate goal is to find a way to 
determine the
extreme eigenvalues (and thereby the dynamics) of networks
by only using averages and local information 
about the network structure. This problem is properly defined
for ensembles of networks and involves two elements:
determination of the ensemble averages and characterization of the
fluctuations across the ensemble. 
Previous studies on ensemble averages have focused
on spectral densities
\cite{Farkas2001,Dorogovtsev2003,Chung2003,taraskin} and
applications \cite{nbw:2006,b:2006,Albert2002,Barahona2002,Nishikawa2003,Motter2005},
while here we focus on the extreme eigenvalues.
In this case, the study of fluctuations
is crucial to assess how well the averages reflect
the properties of individual networks in the ensemble. 
The broader the distributions of extreme eigenvalues across the ensemble, the more limited
the information provided by the averages will be.
It has been suggested recently that the degree distribution 
and other statistical properties are {\it not} sufficient to characterize 
the eigenvalues of random SF networks \cite{Wu}.
Though the spectral properties of these networks are
different from those traditionally considered in random matrix theory 
\cite{dean2006}, as far as we know, extreme eigenvalue
distributions have not been studied for network ensembles and 
their statistical properties remain essentially unknown.

In this Letter, we investigate the {\it averageability} of the extreme
eigenvalues in ensembles of random SF networks. We define a quantity 
to be ensemble averageable if the variance of its probability
distribution goes to zero in the limit of large system size.
We show that the largest eigenvalues of the Laplacian and adjacency 
matrices are determined mainly by the largest degree node
of the network, while the smallest nonzero eigenvalue of the Laplacian depends
on the details of the way in which nodes are connected.
We provide strong evidence that the smallest nonzero 
eigenvalue of the Laplacian and both extreme 
eigenvalues of the normalized Laplacian are ensemble averageable.
That is, as the number of nodes increases, the distributions become increasingly
more peaked and  the averages provide increasingly more accurate information about
the behavior of most networks in the ensemble.
We apply these findings to the study of synchronization and 
epidemic spreading. We show that the physical quantities characterizing 
the dynamics are averageable and properly represented by functions 
of the averages of the extreme eigenvalues. This provides an unambiguous 
spectral characterization of the dynamics in ensembles of large-size networks.  

We focus on undirected random SF networks with the constraints of 
having a single connected component and no self- or multiple links. 
Starting with a graphic degree sequence for $N$ nodes 
generated from a power-law distribution $P_d(k)=c_d k^{-\gamma}$ with $k\geq k_0$, 
where $c_d\simeq (\gamma-1)k_0^{\gamma-1}$ for large $N$, 
we construct an initial network satisfying the given constraints~\cite{kmin}.
Then, we randomize the network topology
using the degree-preserving algorithm of Ref.~\cite{Newman2002}
to implement $(\sum_i k_i)^2$ link-rewirings,
while keeping the constraints by rejecting constraint-breaking rewirings.
The only degree correlations in the network construction are those due to these 
constraints (see, for example, Ref.~\cite{intrinsic}).
We focus on the ensemble of all such networks.
We consider 
three connectivity matrices of broad interest:
the adjacency matrix $\mathrm{A}$, defined as $A_{ij}=1$ if nodes $i$ 
and $j\neq i$ are connected and $A_{ij}=0$ otherwise; 
the Laplacian $\mathrm{L}\equiv\mathrm{D}-\mathrm{A}$ 
and the normalized Laplacian $\widetilde{\mathrm{L}}\equiv\mathrm{D}^{-1}\mathrm{L}$,
where $\mathrm{D}=\mathrm{diag}\{k_1,\ldots ,k_N\}$ is the diagonal matrix of
degrees. For undirected networks, all the eigenvalues of these matrices 
are real. The eigenvalues of $\mathrm{L}$ and $\widetilde{\mathrm{L}}$ 
can be ordered as  $\lambda_1= 0 < \lambda_2 \le \cdots \le \lambda_N$ and 
$\mu_1=0 < \mu_2 \le \cdots \le \mu_N \le 2$, respectively. 
The largest eigenvalue of $\mathrm{A}$ is positive and is denoted 
by $\Lambda_N$.
The nodes are labeled in increasing order of their degrees $k_i$,
such that $k_1\le k_2 \cdots \le k_N$.

\begin{figure}[t]
\includegraphics[width=0.47\textwidth]{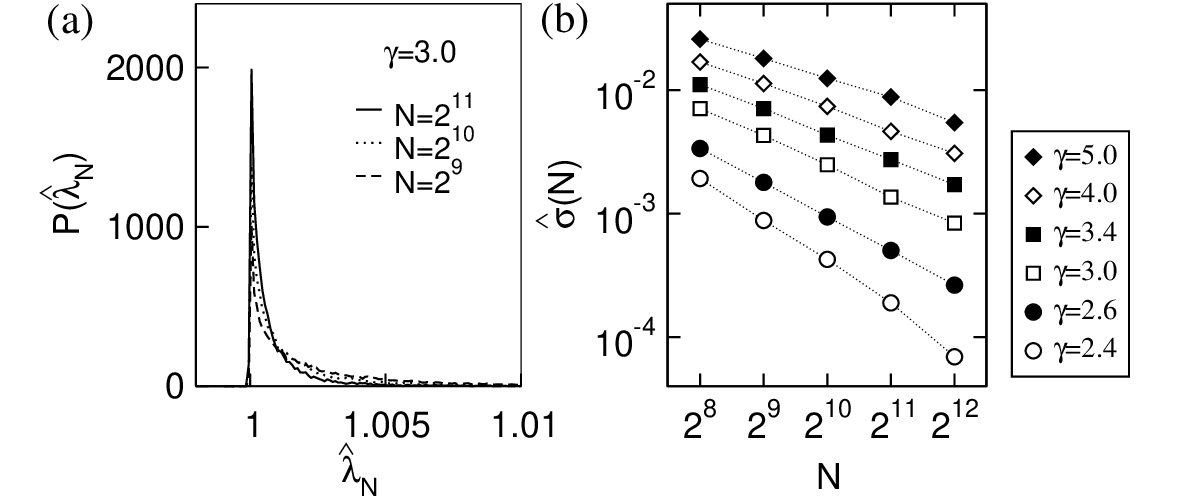}
\caption{
\label{fig1}
Numerical results for (a) the distribution $P(\hat{\lambda}_N)$ 
of $\hat{\lambda}_N \equiv \frac{\lambda_N}{k_N+1}$ 
and (b) the $N$-dependence of the corresponding
standard deviation $\hat{\sigma}$.
All the numerics are obtained from $5000-10^5$ realizations of 
the networks with $k_0=3$.
}
\end{figure}

For the Laplacian $\mathrm{L}$, we estimate the largest eigenvalue $\lambda_N$ 
by using nondegenerate perturbation theory~\cite{perturb}. 
In $\mathrm{L}=\mathrm{D}-\mathrm{A}$, we consider $\mathrm{D}$ as 
an unperturbed matrix and $-\mathrm{A}$ as a perturbation. 
This decomposition leads to the perturbation expansion of $\lambda_N$, 
which up to second order of $\mathrm{A}$ is:   
\begin{equation}
\label{eq:lambdaN}
\lambda_N \simeq k_N - A_{NN} + 
\sum_{j\neq N} \frac{(A_{Nj})^2}{k_N-k_j} \simeq k_N + 1. 
\end{equation}
Here we have used the fact that the second order term can be expanded as 
$\sum_j(A_{Nj})^2(\frac{1}{k_N} + \frac{k_j}{k_N^2} + \cdots) 
= 1 + \frac{k_N^{(1)}}{k_N} + \cdots$, where $k_N^{(1)}$ 
is the average degree of the nearest neighbors of node $N$. 
In uncorrelated SF networks, 
$k_N^{(1)}/k_N \simeq k_N^{-1}\frac{\sum_k k^2P_d(k)}{\sum_k kP_d(k)} \ll 1$ 
for $\gamma>2$ and large $N$~\cite{footnote0}, 
which leads to $\lambda_N\simeq k_N +1$ for large $N$. This
result is quite neat since $\lambda_N$ of any network is lower bounded by $k_N+1$ \cite{bound}.
The same approach can also be used to predict a power-law tail for 
the $\lambda$-density, $\rho(\lambda)\sim \lambda^{-\gamma}$~\cite{dkim2006},
because $\lambda_i \sim k_i$ is still valid for other nondegenerate $k_i$'s 
in the tail of $P_d\sim k^{-\gamma}$. Similarly, we can also obtain
the largest eigenvalue $\Lambda_N$ of the adjacency matrix
by considering the largest
diagonal term of matrix  $\mathrm{A}^2$, given by $k_N$, and regarding the 
off-diagonal elements as a perturbation. Under the approximation
of local tree-like structure, we obtain $\Lambda_N^2 \simeq k_N + k_N^{(1)}-1$,
which provides a second-order correction to the previous result
$\Lambda_N^2 \sim k_N$ \cite{Farkas2001,Dorogovtsev2003,Chung2003}.

Equation~(\ref{eq:lambdaN}) implies that $\lambda_N$ depends on
the specific realization of the degree sequence,
which fluctuates widely across the ensemble.  
For $N$ integers randomly generated from $P_d(k)$, 
the asymptotic form of the probability distribution of the largest one, 
$k_N$, is given by the Fr\'echet distribution
$P(k_N) \simeq c_dNk_N^{-\gamma}\exp[-N(\frac{k_0}{k_N})^{\gamma-1}]$
\cite{Extreme}. 
The average of $k_N$ can be obtained from $P(k_N)$ as  
$\langle k_N \rangle \simeq
k_0N^{\frac{1}{\gamma-1}}\exp[\frac{k_0^{\gamma-1}}
{N^{\gamma-2}}]\Gamma[\frac{\gamma-2}{\gamma-1},\frac{k_0^{\gamma-1}}{N^{\gamma-2}}]$, 
where $\Gamma[a,b]$ denotes the Incomplete Gamma Function.
The standard deviation of $P(k_N)$ increases as 
$\sim N^{\frac{1}{\gamma-1}}$ 
in the same way as $\langle k_N \rangle$ does. This implies that $k_N$, and 
hence $\lambda_N$, are not averageable quantities in this ensemble.

Instead, the corresponding ensemble averageable quantity is 
the \textit{reduced} 
largest eigenvalue $\hat{\lambda}_N\equiv \lambda_N/(k_N+1)$.
While $\hat{\lambda}_N$
may
deviate from the prediction in Eq.~(\ref{eq:lambdaN}),
the numerical calculation confirms that, as $N$ grows,
the distribution of $\hat{\lambda}_N$ becomes extremely peaked,
as shown in Fig.~\ref{fig1}.
This indicates that $\lambda_N$  of large random SF networks is accurately
determined exclusively by $k_N$, which involves local information only.

\begin{figure}[t]
\includegraphics[width=0.47\textwidth]{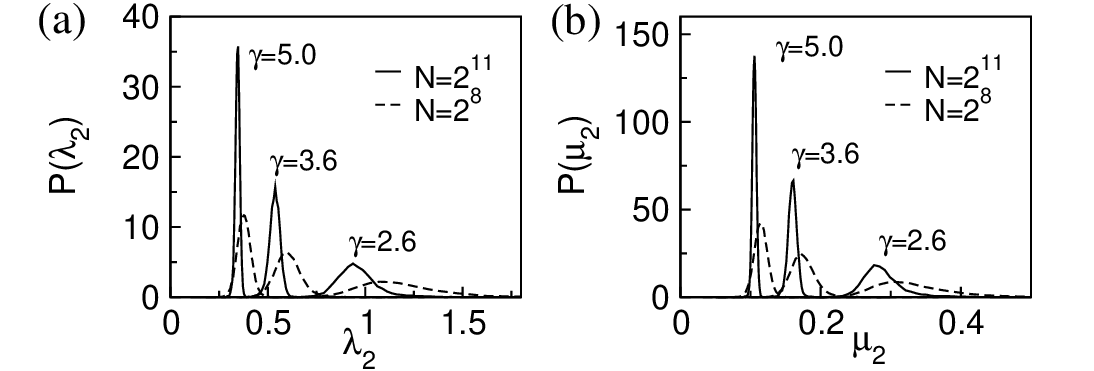}
\caption{
\label{fig2}
Ensemble distributions of (a) $\lambda_2$ and (b) $\mu_2$.
The statistics of $\mu_N$ is not shown because it is well approximated by that
of $2-\mu_2$. 
The unspecified parameters are defined in 
Fig.~\ref{fig1}.
}
\end{figure} 

On the other hand, we find that $\lambda_2$, $\mu_2$ and $\mu_N$ 
are ensemble averageable by themselves. 
As shown in Fig.~\ref{fig2}, these eigenvalues have bell-shaped distributions 
with well-defined averages in the ensemble of SF networks.  
We note that
the statistics of $\mu_N$ is indistinguishable from that of $2-\mu_2$.
More important, we find that $P(\lambda_2)$ and 
$P(\mu_2)$ converge to increasingly peaked distributions as $N$ increases.
We have confirmed this behavior by analyzing the $N$-dependence of the 
standard deviation, which decreases with increasing $N$.
This indicates that the probability of having large deviations 
from the averages decreases to very small values as the size of the system increases.
Therefore, for large $N$, the eigenvalues $\lambda_2$, $\mu_2$ and $\mu_N$ of 
most networks in the ensemble are well represented by the ensemble averages.

To provide approximate bounds for the ensemble averages,
we derive an approximation 
for the extremes 
of the \textit{spectral density} of uncorrelated 
tree-like networks in the thermodynamic limit. For the Laplacian $\mathrm{L}$, 
the spectral density is given by 
$\rho(\lambda)=-\frac{1}{\pi N}\mathrm{Im}\Big\langle\mathrm{Tr}\frac{1}{(\lambda+i0^+)\mathrm{I}-\mathrm{L}}\Big\rangle$
and can be analyzed using a weighted version of the random walk 
method~\cite{Dorogovtsev2003,redner} with a multiplying weight factor 
of $[k_j-(\lambda+i0^+)]^{-1}$.
This leads to
$\rho(\lambda)\simeq \frac{1}{\pi}\mathrm{Im}
\sum_k\frac{P_d(k)}{k-\lambda -i0^+ +kT(\lambda)},$
where $T(\lambda)$ satisfies 
$
T(\lambda) = \frac{1}{\langle k\rangle}\sum_k
\frac{kP_d(k)}{k-\lambda-i0^+ -(k-1)T(\lambda)},
$
with $\langle k\rangle= \sum_k k P_d(k)$. 
To obtain the lower extreme $\lambda^-$
we note that $T$ is complex (so that $\rho(\lambda)>0$)
if $\lambda>\lambda^-$, and $T$ is real (so that $\rho(\lambda)=0$)
if $\lambda<\lambda^-$.
Then, for real $x\equiv\frac{\lambda-T}{1-T}$ such that
$g(x)\equiv \frac{4}{\langle k \rangle}\sum_k \frac{kP_d(k)}{k-x}\le 1$,
we obtain
\begin{equation}
\label{eq:lam2}
\lambda^- \simeq \max_{x}\frac{1}{2}\left[x+1+|x-1|
\sqrt{1-g(x)}\right].
\end{equation}

For the normalized Laplacian $\widetilde{\mathrm{L}}$, it is 
known~\cite{Dorogovtsev2003} that the spectral density is given
by $\rho(\nu) \simeq -\frac{1}{\pi}\mathrm{Im}\frac{1}{\nu-T(\nu)}$
with $T(\nu)$ satisfying 
$
T(\nu)=\frac{1}{\langle k \rangle} \sum_k
\frac{kP_d(k)}{k\nu + i0^+ - (k-1) T(\nu)},
$
where $\nu \equiv 1-\mu$. From the identity $T^*(\nu)=-T(-\nu)$ we obtain
the spectral symmetry $\rho(\nu)\simeq\rho(-\nu)$, which helps explain our
numerical result $\langle\mu_N\rangle \simeq 2 - \langle\mu_2\rangle$
(cf. Fig.~\ref{fig2}). 
We then obtain an approximate expression
for the upper (lower) extreme $\mu^+$ ($\mu^-$)
by using the same argument used to derive Eq.~(\ref{eq:lam2}).
For real $x\equiv T/\nu$, 
\begin{equation}
\label{eq:nu}
|1-\mu^\pm|^2\simeq \min_{0<x<1}\left[\frac{1}{\langle k \rangle x}
\sum_k \frac{kP_d(k)}{k-(k-1)x}\right].
\end{equation}  
If $k_0$ is large,  the r.h.s.\ of Eq.~(\ref{eq:nu}) approaches 
$\sum_k 4(k-1)P_d(k)/(k \langle k \rangle )$ and  
$\mu^\pm \simeq 1\pm 2/\sqrt{\langle k \rangle}$,
which agrees with previous results for densely connected networks 
\cite{Chung2003,dkim2006}.

\begin{figure}[t]
\includegraphics[width=0.47\textwidth]{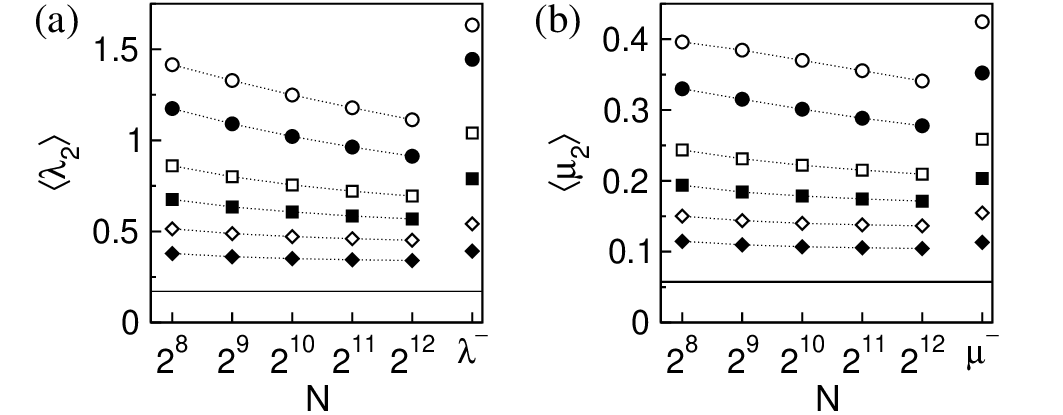}
\caption{
\label{fig3}
Numerical results for the ensemble averages (a) $\langle\lambda_2\rangle$ 
and (b) $\langle\mu_2\rangle$ (dotted lines). We also show
$\lambda^-$ and $\mu^-$ predicted by Eqs. (\ref{eq:lam2}) and 
(\ref{eq:nu}), 
respectively, for the same degree distribution (r.h.s. symbols)
and for $k$-regular graphs with degree $k_0$ (horizontal solid lines). 
The symbols and parameters are the same as in Fig.~\ref{fig1}.
}
\end{figure}

For the large but finite-size sparse networks of our interest, 
the actual ensemble average $\langle\mu_2\rangle$ is expected to fall 
inside of the pseudogap region $(0,\mu^-)$ because of the 
existence of extended tails of $\rho(\mu)$ 
(see Ref.~\cite{Rodgers1988} for the case of homogeneous networks). 
Then, given a degree distribution, $\mu^-$
serves as an approximate upper bound for $\langle\mu_2\rangle$.
On the other hand, because all the networks in the ensemble 
have all degrees $\ge k_0$, the average $\langle\mu_2\rangle$ is expected to be 
lower bounded by the corresponding average
of the ensemble of $k$-regular random graphs with degree $k=k_0$
for all the nodes, which is nonzero for $k_0\ge 3$ \cite{Wu03}.
Thus, we can write $\mu^-_\infty \lesssim \langle\mu_2\rangle \lesssim \mu^-$
and, symmetrically, we have
$\mu^+ \lesssim \langle\mu_N\rangle \lesssim \mu^+_\infty$, 
where $\mu^\pm_\infty$
denotes $\mu^\pm$ at $\gamma=\infty$, representing $k$-regular graphs 
with $P(k)=\delta(k-k_0)$.
For $\lambda_2$, similar arguments lead to 
$\lambda^-_\infty \lesssim \langle\lambda_2\rangle \lesssim \lambda^-$, 
where $\lambda^-_\infty$ is $\lambda^-$ at $\gamma=\infty$. 
As shown in Fig.~\ref{fig3}, the numerical results 
are in good agreement with these predictions.

We now use synchronization and epidemic spreading as example processes
to show how our findings can impact the study of network dynamics. 
In the complete synchronization of identical oscillators, the ability of
an oscillator network to synchronize is measured by the range
of the coupling parameter for which synchronization is stable and is
determined by $R_\lambda\equiv\lambda_N/\lambda_2$ \cite{Barahona2002}. 
If the input signal is normalized to be equal for all the oscillators,
then the same stability condition is determined by $R_\mu\equiv\mu_N/\mu_2$  
\cite{Motter2005}. In epidemic spreading, on the other hand, the epidemic
threshold of the susceptible-infected-susceptible model is determined by 
$1/\Lambda_N$~\cite{WCWF}. These dynamical processes, as well
as many others, are determined by {\it functions} of the extreme eigenvalues. 
To characterize a process in SF networks, in principle one
would have to average the corresponding function over all possible realizations
of the networks or study the process on a case-by-case basis. We have shown,
however, that the extreme eigenvalues are well represented by averages 
combined with local information. 
A practical question then is whether one can approximate the averages
of the functions by functions of the average eigenvalues.

\begin{figure}[t]
\includegraphics[width=0.47\textwidth]{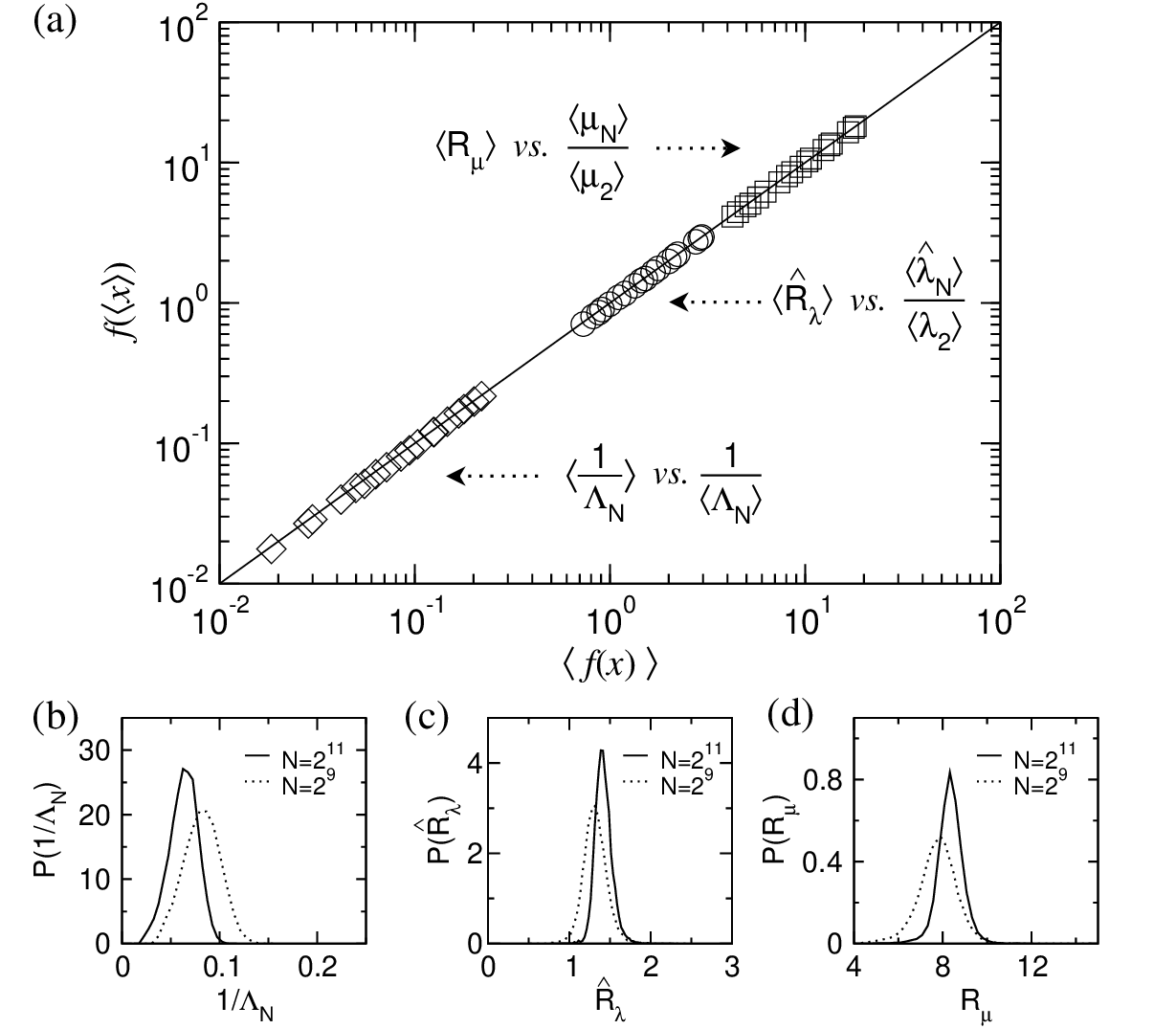}
\caption{
\label{fig4}
(a) $\langle f(x) \rangle$ vs. $f(\langle x \rangle )$ 
for synchronization 
[$\langle R_\mu \rangle$ 
vs. $\frac{\langle\mu_N\rangle}{\langle\mu_2\rangle}$ (squares), $\langle \hat{R}_\lambda \rangle$ 
vs. $\frac{\langle\hat{\lambda}_N\rangle}{\langle\lambda_2\rangle}$ (circles)] and 
epidemics [$\langle\frac{1}{\Lambda_N}\rangle$ vs. 
$\frac{1}{\langle\Lambda_N\rangle}$ (diamonds)]
for networks with $N=2^8-2^{12}$ and $\gamma=2.4-5.0$. 
Panels (b)-(d) show the ensemble distributions for $\gamma=3.0$ and
$N=2^9-2^{11}$.
Here, $\hat{R}_{\lambda}\equiv R_{\lambda}/(k_N+1)$.
The other parameters are the same as in Fig.~\ref{fig1}. 
}
\end{figure}

The average of a function, $\langle f(x) \rangle$, is not necessarily 
equal to the function of the average, $f(\langle x \rangle)$. 
However, from the identity 
$\sum_i f(x_i)/n = f(\sum_i x_i/n)$ for $x_1=x_2=\cdots =x_n$,
one expects that, if the distribution of $x$ goes to 
a $\delta$-like function in the thermodynamic limit, 
then $\langle f(x) \rangle$ approaches $f(\langle x \rangle)$.   
For finite $N$, this can be formalized for locally monotonic functions
by noting that the probability distributions of the function and variable
are related through $P_f(f(x))=P_x(x)/\frac{df(x)}{dx}$. 
If $f(x)$ can be expressed
as a uniformly convergent Taylor series around $x=\langle x \rangle$,
the deviation of $f(\langle x \rangle)$ from  $\langle f(x) \rangle$ 
can be written as
$
\langle f(x) \rangle - f(\langle x \rangle) = \sum_{n=2}^\infty
\frac{1}{n!}f^{(n)}(\langle x \rangle )\langle (x-\langle x\rangle )^n\rangle,
$
where $f^{(n)}(\langle x \rangle)$ denotes the $n$th derivative of $f(x)$ 
at $x=\langle x \rangle$. 
In this case, the central moments $\langle (x-\langle x\rangle )^n \rangle$ 
determine the $N$-dependence of the deviation. If $x$ is averageable, this 
deviation
is expected to decrease as $N$ increases and the central moments decrease.

In Fig.~\ref{fig4}, we show numerically that the averages of the functions  
$\hat{R}_\lambda\equiv R_\lambda/(k_N+1)$, 
$R_\mu$, and $1/\Lambda_N$ are indeed well approximated by the functions of
the average eigenvalues:
\begin{equation}
\label{eq:fvsf}
\langle \hat{R}_\lambda\rangle \simeq 
\frac{\langle\hat{\lambda}_N\rangle}{\langle\lambda_2\rangle},~ 
\langle R_\mu \rangle\simeq\frac{\langle\mu_N\rangle}{\langle\mu_2\rangle},~
\langle \frac{1}{\Lambda_N} \rangle \simeq \frac{1}{\langle \Lambda_N \rangle}.
\end{equation}
In the lower panels of Fig.~\ref{fig4},
we show that the probability distributions
of these functions become increasingly more peaked as $N$ increases, 
which indicates
that the functions themselves are ensemble averageable. 
Note that we have normalized $R_\lambda$ to benefit from the fact that 
$\hat{\lambda}_N$ is averageable.  The function
$R_\lambda$ is broadly distributed in the ensemble but can be estimated for  
individual realizations of the network using 
$R_\lambda \simeq (k_N +1)\langle \hat{R}_\lambda \rangle$.
A similar argument could be used for $1/\Lambda_N$, although in this case
the extreme statistics~\cite{Extreme} of $x=k_N^{-\frac{1}{2}}\sim 1/\Lambda_N$ 
is directly given by the Weibull distribution $P(x) 
\propto
x^{2\gamma-3}\exp(-cNx^{2\gamma-2})$, 
which becomes increasingly peaked as $N$ increases.

The importance of our results is twofold. First, 
despite the rich variety of possible structural configurations 
of individual networks, one can conclude that most networks 
in an ensemble of large SF networks have remarkably similar 
spectral properties.
Second, many network dynamical processes can be described using 
average eigenvalues and local information provided by the degrees, 
which require very few network parameters.  These results 
have broad significance in view of the previous
finding \cite{Wu} that there are networks 
in the SF ensemble with very different extreme
eigenvalues, implying large deviations in the corresponding dynamics.
Our results show that the probabilities
of such large deviations are remarkably small 
and decrease with the increasing size of the networks. 

The averageability of the extreme eigenvalues established in this Letter helps
provide an unambiguous setting for the spectral characterization of dynamical
processes on ensembles of complex networks. For large random SF 
networks, our results show that the eigenvalues $\lambda_2$, $\mu_2$, 
and $\mu_N$ are statistically well characterized by the ensemble averages 
determined by the degree distribution, which is in sharp contrast with the 
conclusions drawn from the study of particular networks \cite{Wu}.
Our conclusion also applies to $\lambda_N$ and $\Lambda_N$ normalized by 
simple functions of the maximum degree. These results provide evidence of 
self-averaging properties reminiscent of the laws of large numbers and 
are likely to remain valid for other ensembles of disordered networks. 

The authors thank David Taylor, Hermann Riecke, and Byungnam Kahng for providing
feedback on the manuscript.


\begin{thebibliography}{}
\bibitem{kmin}
We assume that $k_0\ge 3$ to guarantee the existence of nontrivial
connected configurations for any graphic sequence.
\bibitem{footnote0}
$k_N^{(1)}/k_N \sim k_N^{-1}$ for $\gamma>3$ and
$\sim k_N^{2-\gamma}$ for $2<\gamma<3$.
Thus $k_N \sim N^{\frac{1}{\gamma-1}}$
implies $\lim_{N\to\infty} k_N^{(1)}/k_N = 0$ for $\gamma>2$.
\end{thebibliography}
\end{document}